\documentclass[aps,prb,twocolumn,showpacs,floatfix,superscriptaddress]{revtex4}
\usepackage{amsmath}
\usepackage{amsfonts}
\usepackage{amssymb}
\usepackage{euscript}
\usepackage{bm}
\usepackage{epsfig}
\usepackage{graphicx}

\usepackage{color}

\newcommand{\fgies}[3]{\mbox{\raisebox{#3}
{\epsfig{file=#1,scale=#2,clip=true}}~}}
\begin{document}

\title{Diagrammatic Monte Carlo study  of the Fermi polaron in two dimensions} 
\author{Jonas Vlietinck}
\affiliation{Department of Physics and Astronomy, Ghent University,
Proeftuinstraat 86, 9000 Gent, Belgium}
\author{Jan Ryckebusch}
\affiliation{Department of Physics and Astronomy, Ghent University,
Proeftuinstraat 86, 9000 Gent, Belgium}
\author{Kris Van Houcke}

\affiliation{Department of Physics and Astronomy, Ghent University,
Proeftuinstraat 86, 9000 Gent, Belgium}
\affiliation{Laboratoire de Physique Statistique, Ecole Normale Sup{\' e}rieure, UPMC, Universit{\' e} Paris Diderot, CNRS, 24 rue Lhomond, 75231 Paris Cedex 05, France}

\begin{abstract}
We study the properties of the two-dimensional Fermi polaron model in which
 an impurity attractively interacts with a Fermi sea
of particles  in the zero-range limit. We use a diagrammatic Monte Carlo (DiagMC)
method which allows us to sample a Feynman diagrammatic series to very high
order. The convergence properties of the series and the role of %the
%higher-order
 multiple particle-hole excitations are discussed. We study the
polaron and molecule energy as a function of the coupling strength, revealing a transition from a polaron to a molecule in the ground state. We
find a value for the critical interaction strength which complies with
the experimentally measured one and predictions from variational
methods. For all considered interaction strengths, the polaron $Z$
factor from the full diagrammatic series almost coincides with the 
one-particle-hole result. We also formally link the DiagMC and the variational approaches for the polaron problem at
hand.
\end{abstract}

\maketitle

\section{Introduction}
Experiments with ultracold gases are a powerful tool to investigate
the (thermo)dynamics of quantum many-body systems under controlled
circumstances. 
 With
Feshbach resonances~\cite{feshbach}, for example, one has the ability to tune the
interaction strength.  Optical potentials~\cite{potential} can be exploited to modify
the dimensionality of the studied systems.
The properties of a single impurity that interacts strongly with a background gas,  for example, can  be addressed with ultracold atoms.

The so-called Fermi polaron problem refers to a single spin-down
impurity that is coupled to a non-interacting spin-up Fermi sea (FS).
This problem corresponds to the extreme limit of spin imbalance in a
two-component Fermi gas ~\cite{Partridge06,Shin08,Nascim09} and has
implications on the phase diagram of the strongly spin-polarized Fermi
gas~\cite{Pilati08,pietro,Bertaina}.  At weak attraction, one expects a
``polaron" state~\cite{Chevy06}, in which the impurity is dressed with
density fluctuations of the spin-up Fermi gas.  Recent experiments
 have observed indications of a transition from this polaronic state
to a molecular state (a two-body bound state of the impurity and an
atom of the sea) upon increasing the attraction strength in three
dimensions (3D) ~\cite{polaronMIT} and in two dimensions (2D)
~\cite{kos}. Experimentally, the 2D regime can be accomplished by means
of a transverse trapping potential $V(z)=\frac{1}{2}m\omega_z^2 z^2$
(here, $\omega_z$ is the frequency and $z$ is the transverse direction)
that fulfills the condition $k_B T \ll \epsilon_F \ll \hbar\omega_z$
($T$ is the temperature and $\epsilon_F$ is the Fermi energy of the
FS). When excitations in the $z$ dimension are possible, one reaches the so-called quasi-2D regime ~\cite{jesper2,dyke}.
The purely 2D limit
is reached for $\epsilon_F/\hbar \omega_z \rightarrow 0$ and will be the subject of this paper.

The existence of a polaron-molecule transition in  3D has been predicted with the aid of the diagrammatic
Monte Carlo (DiagMC) method ~\cite{polaron1,polaron2,polaron} and of 
variational methods~\cite{Chevy06,Punk09,Combescot09,Mora}. For the latter, the maximum number of particle-hole (p-h)
excitations of the FS is limited to one or two~\cite{Chevy06,Punk09,Combescot09,Mora}. 
One might naively expect that
the role of quantum fluctuations increases in importance with decreasing dimensionality and that
high-order p-h excitations could become more important in one and two dimensions.  For
the one-dimensional (1D) Fermi polaron the known analytical solution displays no polaron-molecule
transition~\cite{oneDexact}.  
Like for the 3D polaron, the approximate method in
which the truncated Hilbert space contains one p-h and
two p-h excitations of the FS gives results for
the 1D polaron approaching the exact solution
~\cite{oneD1,oneD2}. In 2D, the Fermi polaron properties have  been  
studied with variational wave functions ~\cite{Zollner11,Parish11,Levinsen13}. 
To observe a polaron-molecule transition in 2D it is crucial to include particle-hole excitations in both the polaron
and molecule wave functions~\cite{Parish11}. 
In the limit of weak interactions, the 1p-h and 2p-h variational Ans\"atze for the polaron branch provide similar
results.  Surprisingly, this is also the situation for strong correlations~\cite{Levinsen13}.    

In this work we focus on the 2D Fermi polaron for attractive interactions and
study the role of multiple particle-hole (mp-h) excitations for the
ground-state properties of the system.  The quasiparticle properties
of the polaron are computed with the DiagMC method. This technique
evaluates stochastically to high order a series of Feynman diagrams  for the
one-particle and two-particle self-energies. 
For the details of
the DiagMC method and the adopted method for determining the
ground-state energies from the computed self-energies, we refer to
Refs.~\cite{polaron,polaron2}. In this work we present DiagMC predictions for
the interaction-strength dependence of the polaronic and molecular
ground-state properties in 2D.  We first briefly discuss the model and the diagrammatic method. 
We then discuss the results of the simulations, with particular emphasis on the role of  the mp-h excitations. We also
discuss how variational results for the polaron problem can be obtained within the DiagMC formalism. 

\section{Formalism}\label{sec1}
\par We consider a two-component Fermi gas confined to 2D at
temperature $T=0$. % with a short-ranged interaction between the two
% species. 
Even though we will consider the zero-range interaction in continuous space, we start from a lattice model to avoid
ultraviolet divergences from the onset.
% % Even though our Monte Carlo scheme works directly with the zero-range interaction in continuous space, it is
% convenient to start with a lattice model, thereby elimi- nating ultraviolet divergences at the initial steps of con-
% structing the formalism. The Hamiltonian reads
The corresponding Hamiltonian reads
\begin{multline}
%\begin{left}
\hat{H}  =      \sum_{\mathbf{k}\in \mathcal{B}, \sigma = \uparrow \downarrow} 
\epsilon_{\mathbf{k}  \sigma}  ~ \hat{c}^{\dagger}_{\mathbf{k} \sigma}
\hat{c}^{\phantom{\dagger}}_{\mathbf{k} \sigma}
%  +   \\  \frac{1}{\mathcal{V}} \sum_{\mathbf{k}, \mathbf{k}', \mathbf{q} \in \mathcal{B}}
%V(\mathbf{k} - \mathbf{k}')  ~
%\hat{c}^{\dagger}_{\mathbf{k}+\frac{\mathbf{q}}{2} \uparrow}
%\hat{c}^{\dagger}_{-\mathbf{k}+\frac{\mathbf{q}}{2} \downarrow} 
%\hat{c}^{\phantom{\dagger}}_{-\mathbf{k}'+\frac{\mathbf{q}}{2} \downarrow}
%\hat{c}^{\phantom{\dagger}}_{\mathbf{k}'+\frac{\mathbf{q}}{2} \uparrow} \; .  \\
\\ +  g_{0}    \sum_{\mathbf{r}} b^2 ~ \hat{\Psi}^{\dagger}_{\uparrow} (\mathbf{r})  \hat{\Psi}^{\dagger}_{\downarrow}
(\mathbf{r}) \hat{\Psi}^{\phantom{\dagger}}_{\downarrow}(\mathbf{r})
\hat{\Psi}^{\phantom{\dagger}}_{\uparrow}(\mathbf{r}) \; ,
\label{eq:ham}
%\end{left}
\end{multline}
with  $\hat{\Psi}^{\phantom{\dagger}}_{\sigma}(\mathbf{r})$ and $ \hat{c}^{\phantom{\dagger}}_{\mathbf{k},\sigma}$
being the  operators for
annihilating a spin-$\sigma$ fermion with mass $m_{\sigma}$ and dispersion $\epsilon_{\mathbf{k} \sigma} = k^2 / 2
m_{\sigma}$ in position and momentum space. 
The components of the position  vector $\mathbf{r}$ are integer multiples of the finite lattice spacing $b$. 
Further, $g_0$ is the bare interaction strength. The wave vectors $\mathbf{k}$ are in the first Brillouin zone
$\mathcal{B} = ]-\pi/b, \pi/b]$. The continuum limit is reached for $b \rightarrow 0$.
%The operators $  \hat{c}^{\dagger}_{\mathbf{k} \sigma}$
%$ \left( \hat{c}^{\phantom{\dagger}}_{\mathbf{k} \sigma} \right)$ create (annihilate) fermions
%with momentum $\mathbf{k}$ and spin $\sigma$. 
%The spin-$\sigma$ fermions have
% mass $m_{\sigma}$ and dispersion $\epsilon_{\mathbf{k} \sigma} = k^2 / 2
% m_{\sigma}$, and $\mathcal{V}$ is the area of the system.  
We adopt the convention $\hbar=1$ and consider the mass-balanced case $m_{\uparrow} = m_{\downarrow} = m$. %We model t
%The short-ranged interaction potential is modeled  with a contact $\delta$-interaction,
%$V(\mathbf{r}) = g_0 \delta(\mathbf{r})$, with $g_0$ the bare interaction strength.
%To regularize the ultraviolet divergence caused by the $\delta$-interaction, we first consider fermions on a cubic
lattice with spacing $b$. %The interaction becomes on-site, or $V(\mathbf{r}) = g_0 \delta_{\mathbf{r},\mathbf{0}}/b^3$
% with $\delta$ the Kronecker delta.
We make use of the $T$ matrix~\cite{landau} for a single
spin-$\uparrow$ and spin-$\downarrow$ fermion in vacuum, 
\begin{equation}
 -\frac{1}{g_0} = \frac{1}{\mathcal{V}} \sum_{\mathbf{k}  \in \mathcal{B}}
\frac{1}{\varepsilon_B + \epsilon_{\mathbf{k} \uparrow} + \epsilon_{\mathbf{k}
\downarrow} } \; ,
\label{eq:tvacuum}
 \end{equation}
where $\mathcal{V}$ is the area of the system and $\varepsilon_B$ is the two-body
binding energy [which depends on $m$, $g_0$, and $b$ and $\varepsilon_B(m,g,b)>0$] of a weakly bound state. Such a
state  always exists for an attractive interaction in 2D. With the above relation we 
eliminate the bare interaction strength $g_0$ in favor of the quantity $\varepsilon_B$.  Moreover, the diagrammatic
approach allows us to take the continuum limit $b
\to 0$ and $g_0\to 0^-$ while keeping $\varepsilon_B$ fixed. 
Summing all ladder diagrams gives a partially dressed interaction vertex $\Gamma ^0$:
\begin{equation}
\fgies{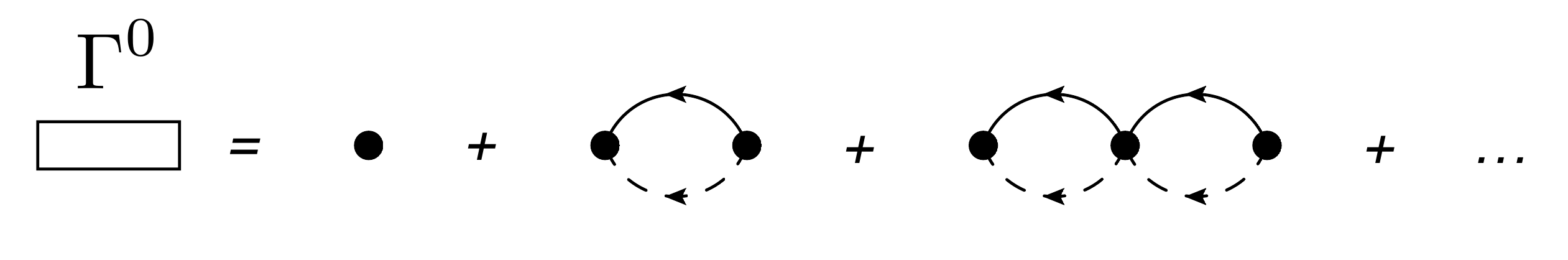}{0.3}{0mm} \;,
\label{eq:Gamma0_diag}
\end{equation}
where the dot represents the bare interaction vertex $g_0$ and the lines represent bare-particle propagators for the
spin-down impurity (dashed lines) and the  spin-up Fermi sea (solid lines).
In momentum-imaginary frequency this graphical representation corresponds to
\begin{equation}
[\Gamma^0(p,i\Omega)]^{-1} = g_0^{-1} - \Pi^0(p,i\Omega) \; ,
\label{eq:bethe}
\end{equation}
 with 
 \begin{eqnarray}
  \Pi^0(p,i\Omega) =  \frac{1}{\mathcal{V}} \sum_{\mathbf{k} \in \mathcal{B}} 
  \frac{H(|\frac{\mathbf{p}}{2}+\mathbf{k}|-k_F)}{i\Omega -  
\epsilon_{\frac{\mathbf{p}}{2}-\mathbf{k} \downarrow}  -
   \epsilon_{\frac{\mathbf{p}}{2}+\mathbf{k} \uparrow}  +\mu + \varepsilon_F} \;
,
\label{eq:Pi0}
 \end{eqnarray}
 with $H(x)$ being the Heaviside step function and $\mu<0$ being a free parameter
representing an energy offset of the impurity dispersion. Further, 
$k_F$ and $\varepsilon_F = \frac {k_F^2} {2m}$ 
are the Fermi momentum and the Fermi energy of the spin-up sea. 
The combination of Eqs.~(\ref{eq:tvacuum}) and (\ref{eq:bethe}) gives

\begin{eqnarray}
\frac{1}{\Gamma^0(p,i\Omega)} = & - & \frac{1}{\mathcal{V}}
\sum_{\mathbf{k} \in \mathcal{B}}\left[\frac{1}{\varepsilon_B +
    \epsilon_{\mathbf{k} \uparrow} + \epsilon_{\mathbf{k} \downarrow}
  } \right. \nonumber \\ &+& \left. \frac{H(|\frac{\mathbf{p}}{2}+\mathbf{k}|-k_F)}{i\Omega -
    \epsilon_{\frac{\mathbf{p}}{2}-\mathbf{k} \downarrow} -
    \epsilon_{\frac{\mathbf{p}}{2}+\mathbf{k} \uparrow} +\mu +
    \varepsilon_F}\right] \; .
\label{eq:renorm}
\end{eqnarray}
The relevant parameter that characterizes the interaction in Equation.~(\ref{eq:renorm}) is 
$\epsilon_B$. Eq.~(\ref{eq:renorm})  is 
well defined in the thermodynamic and  
$b \to 0$ limits. One finds  
\begin{eqnarray}
\frac{1}{\Gamma^0(p,i\Omega)} =   \frac{m}{4\pi}  ~   {\rm ln}\bigg[ 
\frac{2\varepsilon_B}{ -z +  
\sqrt{(z-\epsilon_{\mathbf{p}})^2-4\varepsilon_F\epsilon_{\mathbf{p}}} }    
\bigg]   \;,
\label{eq:gamma0}
\end{eqnarray}
with $z \equiv i\Omega+\mu-\varepsilon_F$. In deriving the above
expression for $\Gamma^0(p,i\Omega)$ we have taken $\mu <
-\varepsilon_F$.  
Since Feynman diagrams for the self-energy will be evaluated in the momentum-imaginary-time representation $(p, \tau)$,
we need to evaluate the Fourier transform
%We now take the Fourier transform of $\Gamma^0(p,i\Omega)$,  
\begin{equation}
\Gamma^0(p, \tau)  =  \frac{1}{2\pi}\int_{-\infty}^{+\infty} ~ d\Omega~ e^{-i
\Omega \tau}~ \Gamma^0(p,i\Omega) \; .
\end{equation}
% which cannot be computed analytically. 
In order to determine the leading behavior of $\Gamma^0(p, \tau)$ for small $\tau$, we introduce the vertex function
$\tilde{\Gamma}^0$, which differs from $\Gamma^0$ by ignoring the 
Fermi surface when integrating out the internal momenta. % Technically
 This amounts to ignoring the Heaviside function in Eq.~(\ref{eq:Pi0}). We obtain
\begin{eqnarray}
 \frac{1}{\tilde{\Gamma}^0(p,i\Omega)} %]^{-1}  
  =   \frac{m}{4\pi}  ~   {\rm ln}\bigg[ - \frac{\varepsilon_B}{i\Omega+\mu+\varepsilon_F -
\frac{\epsilon_{\mathbf{p}}}{2} }    \bigg]   \; .
 \label{eq:gamma0omega}
 \end{eqnarray}
In the $(p,i\Omega)$-representation,
 \begin{eqnarray}
 \frac{1}{\Gamma^0}  - \frac{1}{\tilde{\Gamma}^0}%]^{-1}  
    =   \frac{m}{4\pi}  {\rm ln}\bigg[  \frac{-2(z+2\varepsilon_F)+\epsilon_{\mathbf{p}}}{-z +
\sqrt{(z-\epsilon_{\mathbf{p}})^2-4\varepsilon_F\epsilon_{\mathbf{p}}}} \bigg] \; .
\label{eq:Gammapibar}
 \end{eqnarray}
The $(p,\tau)$ representation of $\tilde{\Gamma}^0$ is
\begin{eqnarray}
 \tilde{\Gamma}^0(p,\tau)  & = &  
-\frac{4\pi\varepsilon_B}{m} e^{-(\frac{\epsilon_{\mathbf{p}}}{2}  -
\varepsilon_F  -\mu )\tau} ~  \bigg[ \int_0^{+\infty} dx
\frac{e^{-x\varepsilon_B\tau}}{\pi^2 +   {\rm ln}^2( x) }  \nonumber  \\ 
& & +  ~ e^{\varepsilon_B \tau} H(\frac{\epsilon_{\mathbf{p}}}{2}  - \varepsilon_F - \varepsilon_B -\mu) \bigg] 
~H(\tau) 
 \label{eq:gam}
 \end{eqnarray}
for  $\mu<  - \varepsilon_F - \varepsilon_B$, ensuring that only $\tau>0$ contributes for  all momenta $p$.
 The integral in Eq.~(\ref{eq:gam}) can be computed numerically, but converges poorly for $\tau \rightarrow 0^+$. 
Under those conditions we make use of the asymptotic behavior: %series expansion:
 \begin{equation}
  \int_0^{+\infty} dx \frac{e^{-x\varepsilon_B\tau}}{\pi^2 +   {\rm ln}^2(
x) } \underset{\tau \rightarrow 0}{\sim}
  \frac{1}{\varepsilon_B\tau}\frac{1}{{\rm ln}^2(\varepsilon_B\tau)} 
\; .
 \end{equation}

To obtain $\Gamma^0(p,\tau)$ we computed numerically the following Fourier transform:
\begin{eqnarray}
 \Gamma^0(p,\tau)-\tilde{\Gamma}^0(p,\tau) & = & \frac{1}{2\pi}\int_{-\infty}^{+\infty}  d\Omega~ e^{-i
\Omega \tau}
\nonumber \\ 
&  \times &  \left[ \Gamma^0 (p,i \Omega) - \tilde{\Gamma}^0 (p,i \Omega) \right]  \; .
\label{eq:eenextra}
\end{eqnarray}
The left-hand side of Eq.~(\ref{eq:eenextra}) can be computed more
easily than $\tilde{\Gamma}^0(p,\tau)$ as it contains no
singularities.  Next, the function $\Gamma^0(p,\tau)$ is obtained as 
$\tilde{\Gamma}^0(p,\tau) + \left[ \Gamma^0 (p,\tau) -\tilde{\Gamma}^0 (p,\tau) \right]$.  Although the functions
$\tilde{\Gamma}^0(p,\tau)$ and $\Gamma^0(p,\tau)$ are extremely sharp
and divergent for $\tau \to 0$, they are integrable.  Special care
should be taken when using these functions in the Monte Carlo code. It
is important to correctly sample very short times, and one needs to
make sure there is no loss of accuracy when keeping track of imaginary
time differences of the $\Gamma^0$ lines in the diagrams.
  \begin{figure}
\includegraphics[width=5cm] {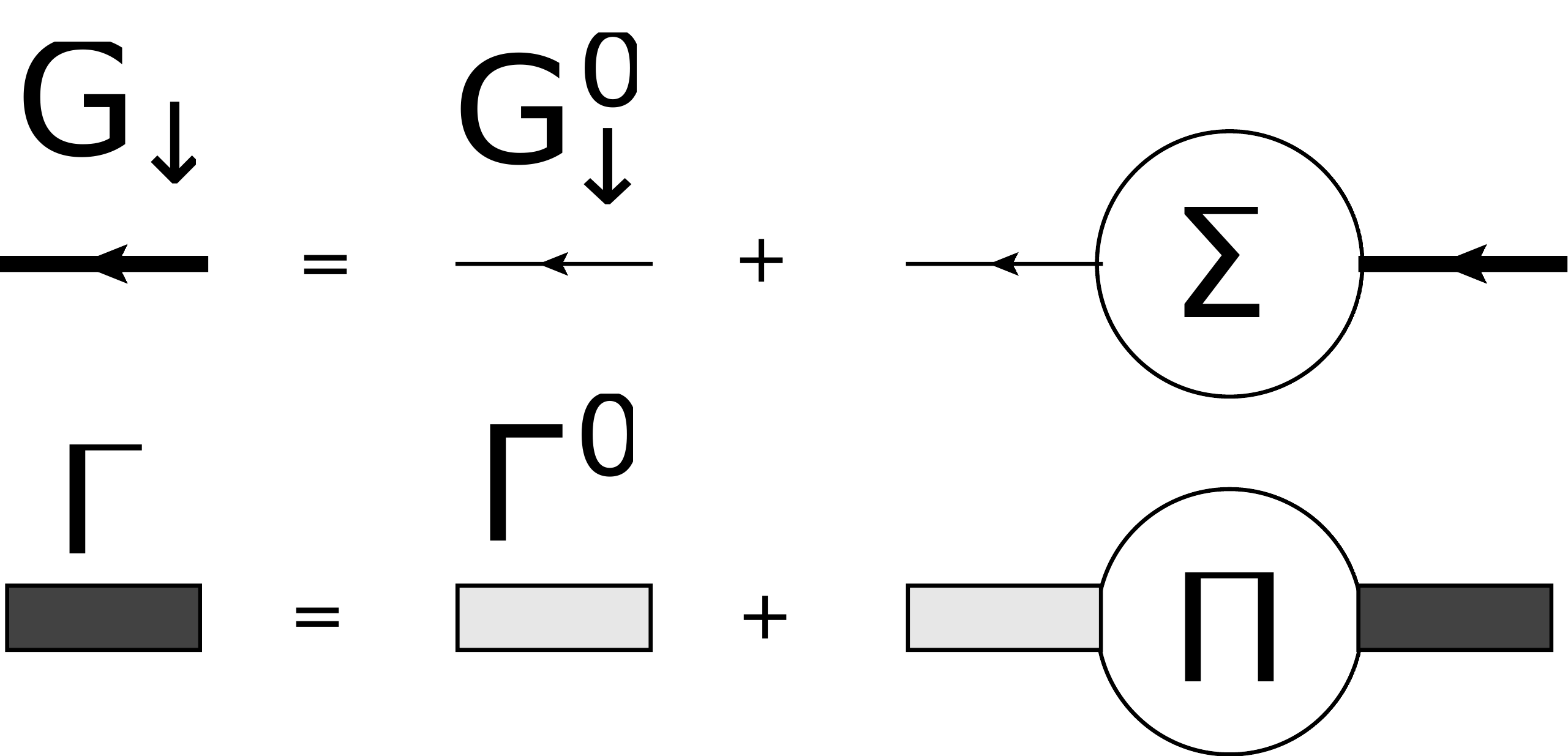}
\caption{\label{fig:dyson} Graphical representation of the Dyson
  equation. The free (dressed) one-body impurity propagator is denoted
  by $G_{\downarrow}^0$ ($G_{\downarrow}$). $\Sigma$ and $\Pi$ are
  the one-body and two-body self-energies, respectively. $\Gamma$
  is the fully dressed interaction, whereas $\Gamma^0$ is the partially
  dressed interaction as shown in Eq.~(\ref{eq:Gamma0_diag}).}
\end{figure}
  Just like for the 3D polaron problem~\cite{polaron1,polaron2,polaron}, we consider a diagrammatic series for the
self-energy built from the free one-body
  propagators for the impurity and the spin-up Fermi sea %particles in the sea. For 
  and from the renormalized
  interaction $\Gamma^0$. We refer to this series as the \emph{bare
  series}, which we evaluate with the DiagMC
  method. 
  The diagram topologies in 2D and 3D are exactly the same.
  The major differences between the diagrammatic-series evaluations in 2D
  and 3D are the renormalized interaction $\Gamma^0(p,\tau)$ and the phase-space volume elements. 
  The one and two-body self-energies are related to the one-particle propagator $G$ and the fully dressed interaction
$\Gamma$ by means of a Dyson equation, as illustrated schematically in Fig.~\ref{fig:dyson}. From the poles of $G$ and
$\Gamma$ we can extract the
polaron and   the molecule
energy, respectively. The fully dressed interaction is closely related to the two-particle propagator ~\cite{polaron}.

  For the 3D
  Fermi polaron problem there are two dominant diagrams  at each given order  that emerge
  next to many diagrams with a much smaller contribution~\cite{polaron}. These dominant diagrams contribute almost
equally but have opposite sign. 
 In 2D, however, the numerical calculations indicate that at a given order the very same two diagrams dominate, but to a
lesser extent; that is, the nondominant diagrams have a larger weight in the final 2D result.
 By \emph{weight} of a given diagram we mean the absolute value of its contribution to the self-energy.
 We note that the sign of  a single diagram at fixed internal and external variables depends only on its topology and
not on the values of the internal and external variables. We stress that this is not true for a Fermi system with two
interacting components with finite density~\cite{vanhoucke12,vanhoucke13}.
 In 2D the total weight of a given order (i.e., the sum of the absolute values of the contributions of diagrams) is
distributed over more diagrams than in 3D.
  Because the sign alternation occurs over a broader distribution of the weights,  we get more 
 statistical noise in sampling the self-energy in 2D compared to 3D. In 3D we can evaluate the diagrammatic series for
the one-body self-energy accurately up to order $12$, whereas in 2D we can reach order $8$. 
 
  In principle, other choices for the propagators (``bare'' versus
  ``dressed'' propagators) are possible, and this was discussed in
  detail for the 3D Fermi polaron in our previous paper
  ~\cite{polaron}.  Replacing the bare propagators by dressed ones reduces the number of diagrams at each given order.
One may expect that this replacement could allow one to reach higher orders.
 For the 3D polaron, however, the most favorable conditions of cancellations between
  the contributions from the various diagrams were met in the bare
  scheme~\cite{polaron}. In the DiagMC framework a
  higher accuracy can be reached under conditions of strong cancellations
  between the various contributions. 
From numerical investigations with various propagators for the 2D Fermi polaron we
could draw similar conclusions as in the 3D studies. Accordingly, all numerical results
for the quasiparticle properties presented below are obtained for 
a series expansion with bare propagators.

%----------------------------------------------------------------------  
    \begin{figure}[ptb]
  \includegraphics[width=\columnwidth] {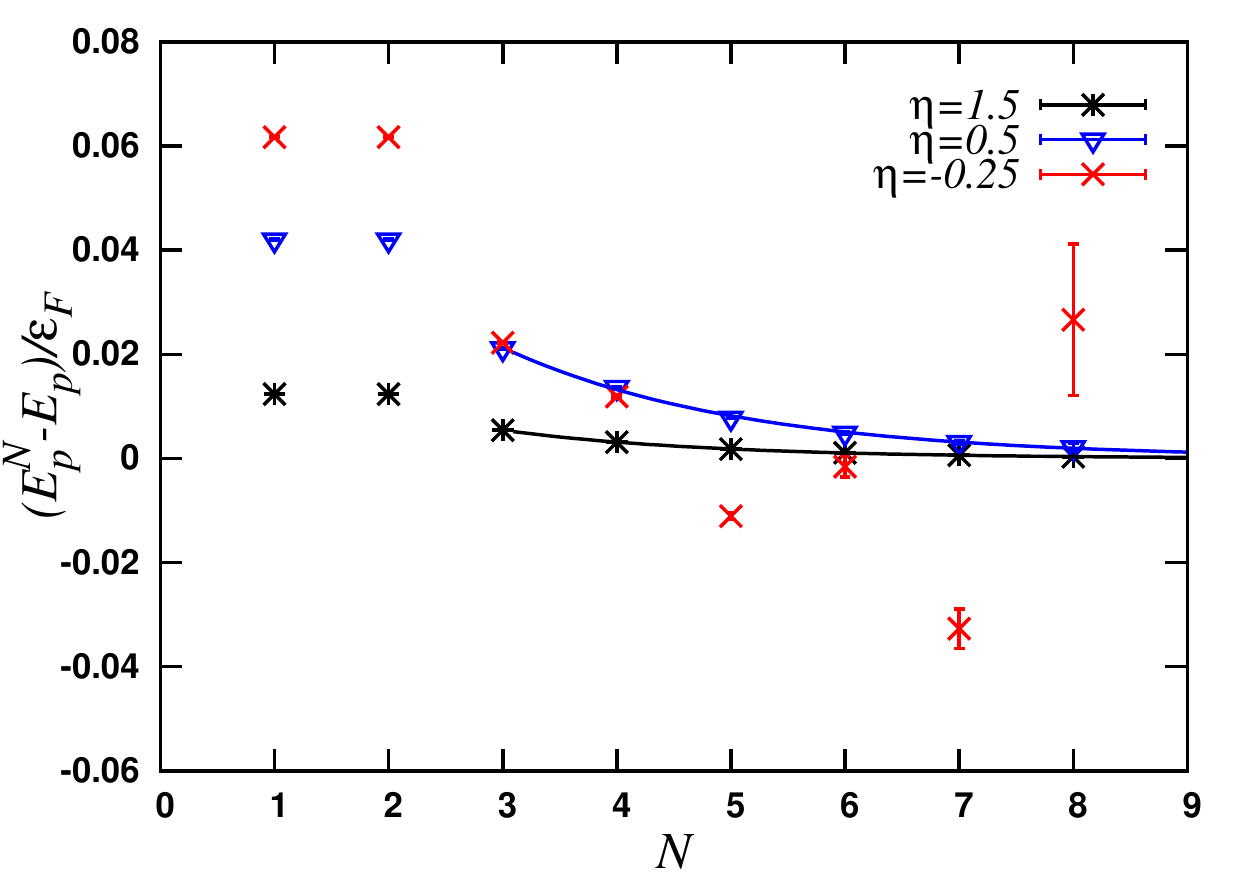}
  \caption{\label{fig:conv}(Color online) Dependence of the polaron energy $E^N_p$ on the  
cutoff diagram order  $N$. $E_p$ is the value obtained after extrapolation to $N\to +\infty$ (and resummation for
$\eta=-0.25$). Results are shown for  $\eta=-0.25 , \eta= 0.5, \eta =1.5$. The lines represent
an
exponential fit. 
}
  \end{figure}
  
%----------------------------------------------------------------------  

\begin{figure}[ptb]
      \includegraphics[width=\columnwidth]{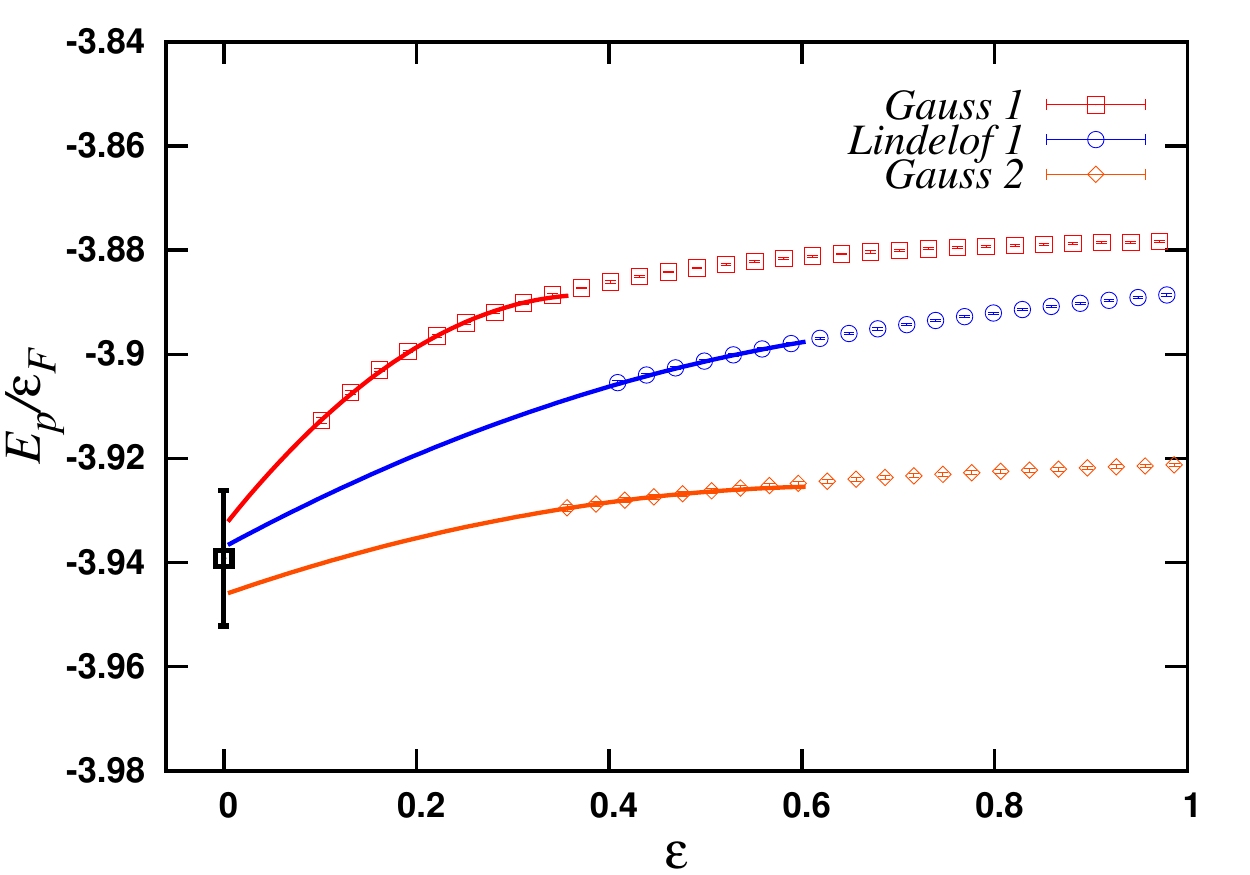}
  \caption{\label{fig:resumpol}(Color online) Abelian resummation of the bare series
    for the one-body self-energy diagrams at $\eta=-0.25$. We evaluate
    $\sigma_\epsilon = \sum_N \sigma^{(N)} e^{-\epsilon \lambda_N}$,
    with $\sigma^N$ being the one-body self-energy for diagram order $N$. We
    use the following functions $\lambda_N$ : (i) Gauss 1: $\lambda_N = (N-1)^2$
    for $N>1$ and $\lambda_N=0$ for $N=1$, (ii) Lindel\"of 1: $\lambda_N $ $= $
    $ (N-1) ~{\log}(N-1)$ for $N> 2$ and $\lambda_N=0$ for $N\le 2$, and (iii) Gauss 2: $\lambda_N = (N-3)^2$ for $N>
3$ and $\lambda_N=0$ for $N\le 3$. The polaron energy $E_{p}/\epsilon_F$ is extracted in the limit
    $\epsilon$ $=$ $0^+$ for various choices of $\lambda_N$.}
\label{resumplot}
  \end{figure}

\begin{figure}[ptb]
  \includegraphics[width=\columnwidth] {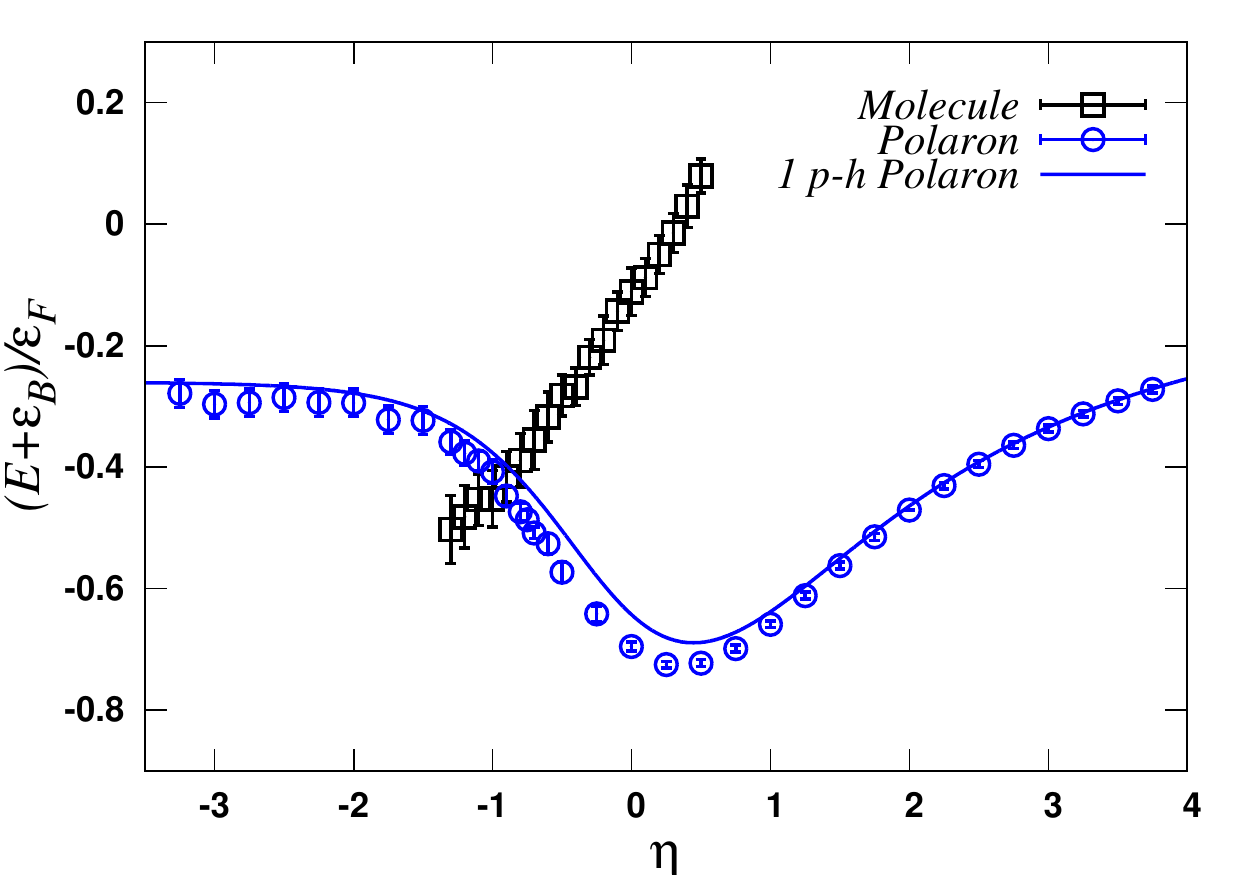}
  \caption{\label{fig:energies}(Color online) Polaron and molecule ground-state
    energies $E$ in units of the Fermi energy $\varepsilon_F$ as a function of  $\eta$. 
    The
    momentum of the impurity is equal to zero.  Energies are shifted by
    the two-body binding energy $\varepsilon_B/\varepsilon_F = 2 e^{-2\eta}$ to magnify the details.  The solid
    line is the DiagMC result for $N=1$.  The symbols are the result of
    the full DiagMC calculations (including diagrams up to order 8).}
  \end{figure}

To characterize the magnitude of the interaction strength we use the dimensionless parameter $\eta \equiv \text{ln}[k_F
a_{2D}] =
\text{ln}[2\varepsilon_F / \varepsilon_B]/2$. Here, $a_{2D}>0$ is the 2D scattering length, related to the dimer binding
energy by $\varepsilon_B=1/(2m_ra_{2D}^2)$  with $m_r = m_{\uparrow}m_{\downarrow}/(m_\uparrow+m_{\downarrow})$ being
the  reduced mass.
The BCS regime corresponds to $\eta \gg 1$ while the Bose-Einstein condensate (BEC) regime corresponds to $\eta \ll -1$.
 The system is perturbative  in the regimes $|\eta| \gg 1$,
 while the strongly correlated regime corresponds to $|\eta| \lesssim 1$.~\cite{bloom}
In the weak-coupling regime [small interaction
strengths $g_0$ in the Hamiltonian of Eq.~(\ref{eq:ham}) or large positive $\eta$ in the zero-range limit], we find
that the one-body
and the two-body self-energy $\Sigma$ and $\Pi$ 
   converge absolutely as a function of the maximum diagram order.  This is
demonstrated in Fig.~\ref{fig:conv} for  $\eta=1.5$, where the polaron energy $E^N_p$
converges exponentially as a function of the cutoff diagram order $N$. Similar convergence is also found for the
molecule energy. Under conditions of convergence with
diagram order, extrapolation to order infinity can be done in a
trivial way.  
Similar convergence is also seen for $\eta=0.5$.
In the strongly correlated regime the series starts
oscillating with order when $\eta \lesssim 0$, and the oscillations get stronger the deeper we go into the BEC regime.
The oscillations in the extracted polaron energy are illustrated in
Fig.~\ref{fig:conv} for $\eta=-0.25$. 
To obtain meaningful results we rely on Abelian resummation techniques ~\cite{polaron,vanhoucke12}. We evaluate the
series 
$\sigma_\epsilon = \sum_N \sigma^{(N)} e^{-\epsilon \lambda_N}$,
    with $\sigma^N$ being the one-body self-energy for diagram order $N$ and $\lambda_N$ being a function that depends
on  the 
diagram order $N$. For each $\epsilon$ the polaron energy $E_p$  is calculated from $\sigma_\epsilon$ and an
extrapolation is done by taking the limit $\epsilon \rightarrow 0$. The whole procedure is illustrated in Fig.~
\ref{fig:resumpol}. To estimate the systematic
error of the extrapolation procedure, different resummation functions $\lambda_N$
are used. As becomes clear from Fig.~\ref{fig:resumpol} the whole resummation procedure is a stable one and induces
uncertainties on the extracted energies of the order of a few percent. 
All the results of Fig.~\ref{fig:energies} are obtained with the Abelian resummation technique. The stronger the
coupling constant is the larger the size of the error attributed to the resummation. An accurate extrapolation to
infinite diagram order could be achieved for all values of $\eta$. 

%----------------------------------------------------------------------

  %----------------------------------------------------------------------

%----------------------------------------------------------------------
%----------------------------------------------------------------------

\section{Results and discussion}\label{sec1}
In Fig.~\ref{fig:energies}, polaron and molecule energies are displayed for a wide range of the parameter $\eta$.  
DiagMC results include all diagrams  up to order 8 and 
extrapolation to the infinite diagram order.  In the region $\eta \lesssim 0$ a small discrepancy (of the order of
$0.1\%$ of the ground-state energy) is found with the 
variational results~\cite{Levinsen13} of Parish and Levinsen based on the wave-function Ansatz up to 2p-h
excitations.  
Clearly, a
phase transition appears at the critical value $\eta_c=-0.95 \pm 0.15$. 
A variational result which includes 2p-h excitations for the polaron and 1p-h excitations for
the molecule, gives $\eta=-0.97$.~\cite{Levinsen13} Both mentioned calculations are in agreement with
the experimental result  $\eta=-0.88(0.20)$~\cite{kos}.

The DiagMC method allows one to include a large 
number of particle-hole excitations that dress the impurity. 
Truncation of the Hilbert space to a maximum number of p-h pairs can nonetheless be achieved within the DiagMC approach.
This allows one to arrive at the variational formulation. 
Previous variational studies using a wave function Ansatz up to 1p-h or 2p-h excitations showed that these
truncations give remarkably accurate results~\cite{Combescot}.

To understand why the truncation is possible within a Feynman diagrammatic approach for the self-energy, we first remark that a variational approach is easily established within a path-integral formalism. 
Path integrals with continuous imaginary time, for example, are based on an expansion of the evolution operator, %. For the partition function one gets
\begin{eqnarray}
e^{-\beta\hat{K}}
  & = & e^{-\beta\hat{K}_0} \big(      1 -  \int_0^{\beta} d\tau_1 K_1(\tau_1)  \nonumber \\  & + &  \int_{0}^{\beta} d\tau_1 \int_{0}^{\tau_1} d\tau_2  ~ \hat{K}_1(\tau_1) \hat{K}_1(\tau_2)  - \ldots     \big)  \; ,
\label{eq:perturb}
\end{eqnarray}
where $\hat{K} = \hat{H}-\mu\hat{N}=\hat{K}_0+\hat{K}_1 -\mu\hat{N}$, with $ \left[ \hat{K}_0 , \hat{K}_1 \right] \ne 0$.  The operator $\hat{K}_1(\tau) = e^{\hat{K}_0\tau} \hat{K}_1 e^{-\hat{K}_0 \tau}$, which defines the series expansion, is expressed in the interaction picture.
Further, $\beta=1/k_BT$, with $k_B$ being Boltzmann's constant and $T$ being temperature, $\hat{H}$ is the Hamiltonian,
$\hat{N}$  the number operator, and $\mu$ is the chemical potential.
The imaginary-time evolution operator in Eq.~(\ref{eq:perturb}) can be used as a ground-state projection operator: for sufficiently long imaginary time $\beta$ the excited-state components of a trial state are exponentially suppressed. 
One typically evaluates all the terms in the expansion equation~(\ref{eq:perturb}) in the eigenbasis of $\hat{K}_0$.
This procedure forms the basis of path-integral Monte Carlo simulation of lattice models, where $\hat{K}_1$ is usually 
the kinetic energy term~\cite{worm}. A discretized time version is used in path-integral Monte Carlo methods in
continuous space~\cite{ceperley,boninsegni}. 
Either way, the contributions to the path integral have the direct physical interpretation of a time history  of the many-particle system. 
At each instant of time, one can constrain the accessible states of the Hilbert space, in line with what is done in a variational approach. 
Within the standard Feynman diagrammatic formalism for Green's functions, however, this truncation of the Hilbert space is not easy to accomplish for an arbitrary system, as one expands in powers of the two-body interaction term of the Hamiltonian. This will be explained in the next paragraph.  

It turns out to be formally easier to start from finite $T$ and to take the $\beta \to \infty$ limit in the end. 
For a many-fermion system, the finite-temperature Green's function in position and imaginary-time representation
$(\mathbf{x},\tau)$ is defined as 
\begin{equation}
G_{\alpha\sigma}(\mathbf{x}, \tau) = - \frac{{\rm Tr}[e^{-\beta \hat{K}}  T_\tau  \hat{\psi}^{\phantom{\dagger}}_{H \alpha}(\mathbf{x},\tau)  \hat{\psi}^{\dagger}_{H \sigma}(\mathbf{x},0)]}{{\rm Tr}[e^{-\beta \hat{K}}]} \; ,
\label{eq:G}
\end{equation}
with $\alpha$ and $\sigma$ denoting an appropriate set of quantum numbers  (such as spin)  and $T_{\tau}$ being the
time-ordering operator.
The field operator in the Heisenberg picture $ \hat{\psi}^{\phantom{\dagger}}_{H \alpha}(\mathbf{x},\tau) = 
e^{\hat{K}\tau} \hat{\psi}^{\phantom{\dagger}}_{\alpha}(\mathbf{x})   e^{-\hat{K}\tau}$ annihilates a fermion in state
$\alpha$ at position $\mathbf{x}$ and time $\tau$. To arrive at the Feynman diagrammatic expansion, one makes a
perturbative expansion for the evolution operator $e^{-\beta \hat{K}}$   in both the numerator \emph{and} the
denominator of Eq.~(\ref{eq:G}) (the finite $T$ ensures that both exist). The expansion of the partition function $Z$ in
the denominator 
can be represented graphically  by the series of all fully closed diagrams (connected and disconnected). 
When $\beta$ approaches $+\infty$, the denominator is proportional to  $\langle \Psi_0 | \Psi_0 \rangle$ ($|\Psi_0\rangle$ is the ground state of the  interacting many-body system), and the disconnected diagrams correspond to all possible vacuum fluctuations of the system at hand. 
The expansion in the numerator factorizes into an expansion of connected diagrams for $G_{\alpha\sigma}$ and
disconnected diagrams for $Z$. So the sum of disconnected diagrams drops out, as expected for an intensive quantity like
$G_{\alpha\sigma}(\mathbf{x}, \tau)$.
It is exactly this factorization that prevents one from truncating the Hilbert space at any instant of time in the
evolution. In other words, variational calculations based on Feynman diagrams for the self-energy are generally not
feasible. 

In the polaron problem vacuum fluctuations are absent since $|\Psi_0\rangle$ corresponds to the spin-down vacuum and a
non-interacting spin-up Fermi sea. In other words, the vacuum cannot be polarized in the absence of an impurity. 
As a consequence, we face a situation similar to the path integral with a direct physical interpretation of the time history of the impurity. This peculiar feature allows one to restrict the Hilbert space at each given time.  
If we allow at most 1p-h excitations at each instant of time, only one diagram survives: the lowest-order self-energy
diagram built from $\Gamma^0$ and the free spin-up single-particle propagator $G^0_{\uparrow}$. The equivalence between
this diagram and the 1p-h variational approach had already been pointed out in Ref.~\cite{oneD1}. An $n$p-h variational
approach is achieved by allowing at most $n$ backward spin-up lines at each step in the  imaginary-time evolution.

For large $\eta$ it
is obvious from Fig.~\ref{fig:energies} that the polaron energy from
the full series expansion becomes equal to the 1p-h 
result. Even for stronger interactions (smaller $\eta$) the
first-order results remain close to the full DiagMC
one. Within the statistical accuracy of the numerical calculations, convergence for the one-body self-energy is already reached after
inclusion of 2p-h excitations.
 Indeed, for all values of $\eta$, we find agreement between our 2p-h variational DiagMC approach and the full DiagMC approach within statistical error bars. 
For the molecular branch, we
retrieve the result for the two-body self-energy from the full series
expansion after including 1p-h 
 excitations. For the
3D Fermi-polaron a similar conclusion was drawn. Also in 3D, the first-order result 
is a very good approximation ~\cite{polaron}. Going up to 2p-h pairs gives a perfect agreement with full DiagMC
results. 
From the above considerations it follows, however, that the diagrammatic truncations which provide good results for the polaron problem may not be appropriate for the more complex many-body problem with comparable densities for both components.

\begin{figure}[ptb]
  \includegraphics[width=\columnwidth] {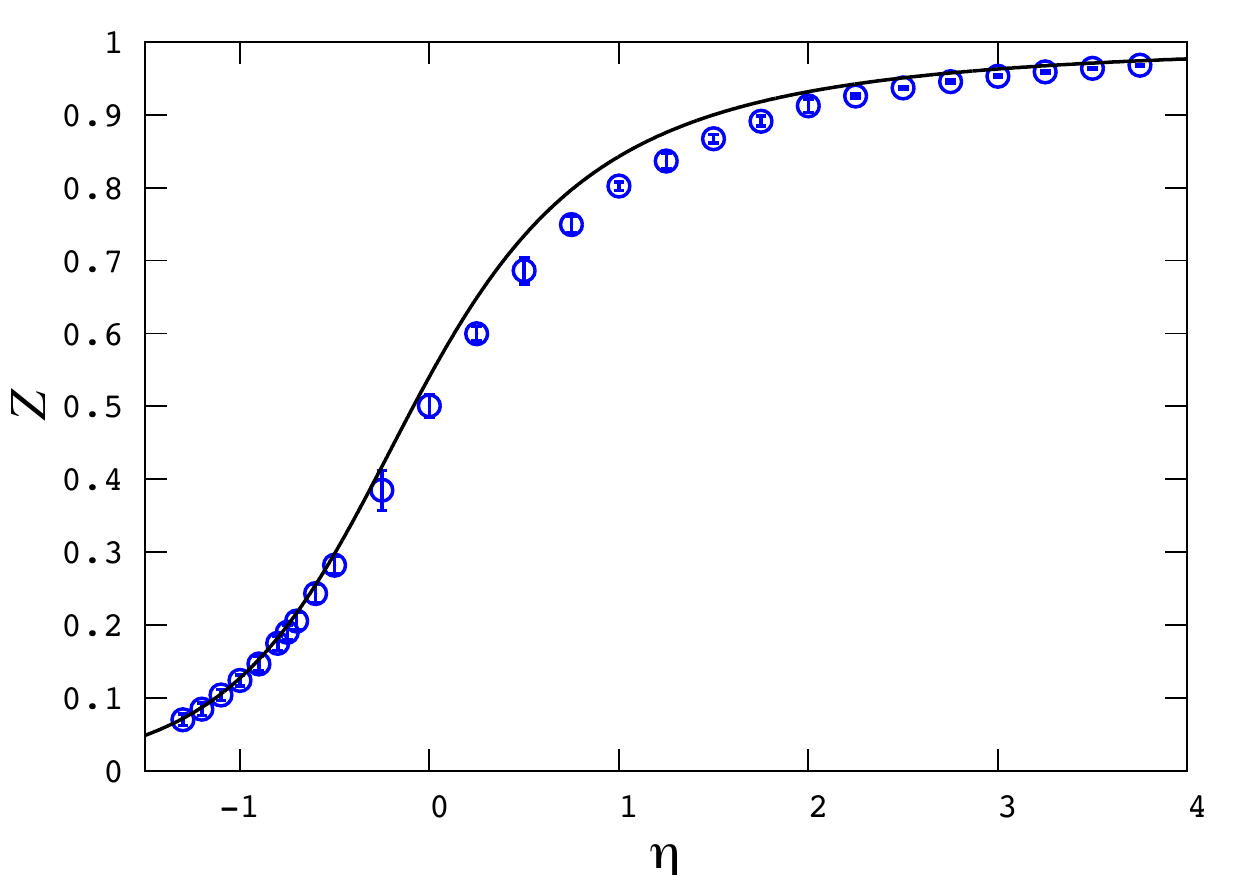}
  \caption{\label{fig:Zpol} (Color online)  The quasiparticle residue $Z$ of the polaron as a function of $\eta$.
The solid line represents  
  the 1p-h result ($N=1$ diagram). }
  \end{figure}

 The quasiparticle residue or $Z$ factor of the polaron gives the
 overlap of the noninteracting wave function and the fully
 interacting one. This overlap is very small for a molecular ground state of the fully interacting
system~\cite{polaron}. The $Z$ factor as a function of
 $\eta$ is shown in Fig.~\ref{fig:Zpol}.  Note that the polaron
 $Z$ factor does not vanish in the region $\eta \lesssim -1$ where the
 ground state is a molecule. The $Z$ factor is, however, still
 meaningful since the polaron is a well-defined (metastable) excited
 state of the 2D system.  Again, the first-order result gives a good
 approximation to the full result. The measured $Z$ factor for the 3D
 situation has been reported in Ref.~\cite{pietro,polaronMIT}. The 2D experimental data are reported in Ref.~\cite{kos},
and
the $\eta$ dependence of the quasiparticle weight $Z$ is presented in arbitrary units. We reproduce the observation that
$Z$ strongly increases between $\eta_c \lesssim \eta \lesssim 1$ and saturates to a certain value for $\eta>1$.

\section{Conclusion}\label{sec1}
Summarizing, we have developed a framework to study with the DiagMC
method the ground-state properties of the 2D Fermi polaron for
attractive interactions.  We have shown that the framework allows one
to select an arbitrary number of $n$p-h excitations of the FS, thereby
making a connection with typical variational approaches which are
confined to $n$=1 and $n$=2. We have studied the quasiparticle properties
of the ground state for a wide range of interaction strengths. A phase
transition between the polaron and molecule states is found at
interaction strengths compatible with experimental values and
with variational predictions. To a remarkable degree, it is observed that
for all interaction strengths the full DiagMC results (which include all
$n$p-h excitations) for the ground-state properties can be reasonably
approximated by n=1 truncations.  
In a n=2 truncation 
scheme the full result is already reached within the %statistical
error bars. 
This lends support for variational approaches to the
low-dimensional polaron problem, for which one could have naively expected
a large sensitivity to quantum fluctuations.

%---------------------------------------------------------------------- 
%----------------------------------------------------------------------  
\section*{Acknowledgments}

% \begin{Acknowledgement}
This work is supported by the Fund for Scientific research - Flanders.
We would like to thank 
C. Lobo,
N. Prokof'ev, 
B. Svistunov, and 
F. Werner
for  helpful discussions and suggestions, and we thank
% We thank 
J. Levinsen and M. Parish for sending us their data. 
% \end{Acknowledgement}

\end{document}